\documentclass[12pt]{article}
\usepackage{amsmath}
\usepackage{amsfonts}
\usepackage{amssymb}
\usepackage{url}

\begin{document}       

\title{Foreword to\\\textbf{A Computable Universe}\\Understanding Computation \& Exploring Nature As Computation\thanks{Published in H. Zenil (ed), \emph{A Computable Universe: Understanding Computation \& Exploring Nature As Computation}, World Scientific, 2012. Footnotes to names are pointers to the chapters in the volume for which this foreword was written.}} \author{Roger Penrose\\Mathematical Institute, University of Oxford, UK}

\date{}

\maketitle

\bigskip \bigskip

I am most honoured to have the privilege to present the Foreword to this fascinating and wonderfully varied collection of contributions, concerning the nature of computation and of its deep connection with the operation of those basic laws, known or yet unknown, governing the universe in which we live. Fundamentally deep questions are indeed being grappled with here, and the fact that we find so many different viewpoints is something to be expected, since, in truth, we know little about the foundational nature and origins of these basic laws, despite the immense precision that we so often find revealed in them. Accordingly, it is not surprising that within the viewpoints expressed here is some unabashed speculation, occasionally bordering on just partially justified guesswork, while elsewhere we find a good deal of precise reasoning, some in the form of rigorous mathematical theorems. Both of these are as should be, for without some inspired guesswork we cannot have new ideas as to where look in order to make genuinely new progress, and without precise mathematical reasoning, no less than in precise observation, we cannot know when we are right---or, more usually, when we are wrong.

The year of the publication of this book, 2012, is particularly apposite, in being the centenary year of Alan Turing, whose theoretical analysis of the notion of ``computing machine'', together with his wartime work in deciphering Nazi codes, has had a huge impact on the enormous development of electronic computers, and on the consequent influence that these devices have had on our lives and on the way that we think about ourselves. This impact is particularly evident with the application of computer technology to the implications of known physical laws, whether they be at the basic foundational level, or at a larger level such as with fluid mechanics or thermodynamics where averages over huge numbers of elementary constituent particles again lead to comparatively simple dynamical equations. I should here remark that from time to time it has even been suggested that, in some sense, the ``laws'' that we appear to find in the way that the world works are \textit{all} of this statistical character, and that, at root, there are ``no'' basic underlying physical laws (e.g. Wheeler's ``law without law'' \cite{wheeler2}, Sakharov's ideas of ``induced gravity'' \cite{sakharov}, etc., and we find this general type of view expressed also in this volume also\footnote{see Calude.\index{Calude, C.}}). However, I find it hard to see that such a viewpoint can have much chance of yielding anything like the enormously precise non-statistical dynamics \cite{schrodinger} and great mathematical sophistication that we find in so much of 20th century physics. This point aside, we find that in reasonably favourable circumstances, computer simulations can lead to hugely impressive imitations of reality, and the resulting visual representations may be almost indistinguishable from the real thing, a fact that is frequently made use of in realistic special effects in films, as much as in serious scientific presentations. When we need precision in particular implications of such equations, we may run into the difficult issues presented by chaotic behaviour, whereby the dependence on initial conditions becomes exponentially sensitive. In such cases there is an effective randomness in the evolved behaviour. Nevertheless, the computational simulations will still lead to outcomes that would be physically allowable, and in this sense provide results consistent with the behaviour of reality.

Computational simulations can have great importance in many areas other than physics, such as with the spread of epidemics, or with economics (where the mathematical ideas of game theory can play an important role),\footnote{Velupillai.\index{Velupillai, K.}} but I shall here be concerned with \textit{physical} systems, specifically. The impressiveness of computational simulations is often most evident when it is simply 17th century Newtonian mechanics that is involved, in its enormously varied different manifestations. The implications of Newtonian dynamical laws can be extensively computed in the modelling of physical systems, even where there may be huge numbers of constituent particles, such as atoms in a simplified gas, or particle-like ingredients, such as stars in globular clusters or even in entire galaxies. It may be remarked that computational simulations are normally done in a time sense where the future behaviour is deduced from an input which is taken to be in the past. In principle, one could also perform calculations in the reverse ``teleological'' direction, because of the time-reversibility of the basic Newtonian laws.\footnote{Beavers.\index{Beavers, A.}} However, because of the second law of thermodynamics, whereby the entropy (or ``randomness'') of a physical system increases with time in the natural world, such reverse-time calculations tend to be untrustworthy.

When Newtonian laws are supplemented by the Maxwell-Lorentz equations, governing the behaviour of electromagnetic fields and their interactions with charged material particles, then the scope of physical processes that can be accurately simulated by computational procedures is greatly increased, such as with phenomena involving the behaviour of visible light, or with devices concerned with microwaves or radio propagation, or in modelling the vast galactic plasma clouds involving the mixed flows of electrons and protons in space, which can indeed be computationally simulated with considerable confidence.

This latter kind of simulation requires that those physical equations be used, that correctly come from the requirements of special relativity, where Einstein's viewpoint concerning the relativity of motion and of the passage of time are incorporated. Einstein's special relativity encompassed, encapsulated, and superseded the earlier ideas of FitzGerald, Lorentz, Poincar\'e and others, but even Einstein's own viewpoint needed to be reformulated and made more satisfactory by the radical change of perspective introduced by Minkowski, who showed how the ideas of special relativity come together in the natural geometrical framework of 4-dimensional space-time. When it comes to Einstein's \textit{general} relativity, in which Minkowski's 4-geometry is fundamentally modified to become \textit{curved}, in order that gravitational phenomena can be incorporated, we find that simulations of gravitational systems can be made to even greater precision than was possible with Newtonian theory. The precision of planetary motions in our Solar System is now at such a level that Newton's 17th century theory is no longer sufficient, and Einstein's 20th century theory is needed. This is true even for the operation of the global positioning systems that are now in common use, which would be useless but for the corrections to Newtonian theory that general relativity provides. Indeed, perhaps the most accurately confirmed theoretical simulations ever performed, namely the tracking of double neutron-star motions, where not only the standard general-relativistic corrections (perihelion advance, rotational frame-dragging effects, etc.) to Newtonian orbital motion need to be taken into account, but also the energy-removing effects of gravitational waves (ripples of space-time curvature) emanating from the system can be theoretically calculated, and are found to agree with the observed motions to an unprecedented precision.

The other major revolution in basic physical theory that the 20th century brought was, of course, \textit{quantum mechanics}---which needs to be considered in conjunction with its generalization to quantum \textit{field} theory, this being required when the effects of special relativity have to be taken into account together with quantum principles. It is clear from many of the articles in this volume, that quantum theory is (rightly) considered to be of fundamental importance, when it comes to the investigation of the basic underlying operations of the physical universe and their relation to computation. There are many reasons for this, an obvious one being that quantum processes are undoubtedly fundamental to the behaviour of the tiniest-scale ingredients of our universe, and also to many features of the collective behaviour of many-particle systems, these having a characteristically quantum-mechanical nature such as quantum entanglement, superconductivity, Bose-Einstein condensation, etc. However, there is another basic feature of quantum mechanics that may be counted as a reason for regarding this scheme of things as being more friendly to the notion of computation than was classical mechanics, namely that there is a basic \textit{discreteness} that quantum mechanics introduces into physical theory. It seems that in the early days of the theory, much was made of this discreteness, with its implied hope of a ``granular'' nature underlying the operation of the physical world. A hope had been expressed\cite{russell,schrodinger} that somehow the domination of physical theory by the ideas of \textit{continuity} and \textit{differentiability}---which go hand-in-hand with the pervasive use of the real-number system---might have at last been broken, via the introduction of quantum mechanics. Accordingly, it was hoped that the ideas of discreteness and combinatorics might soon be seen to become the dominant driving force underlying the operation of our universe, rather than the continuity and differentiability that classical physics had depended upon for so many centuries. A discrete universe is indeed much more in harmony with current ideas of computation than is a continuous one, and many of the articles in this volume\footnote{Bolognesi\index{Bolognesi, T.}, Chaitin\index{Chaitin, G.}, Wolfram\index{Wolfram, S.}, Fredkin\index{Fredkin, Ed}, and Zenil\index{Zenil, H.}.} argue powerfully from this perspective, and particularly in the context of \textit{cellular automata}\footnote{Mart\'inez\index{Mart\'inez, G.J.}, Margenstern\index{Margenstern, M.}, Sutner\index{Sutner, K.}, Wiedermann\index{Wiedermann, J.} and Zuse\index{Zuse, K.}.}.

The very notion of ``computability'' that arose from the early 20th century work of various logicians G\"{o}del, Church\index{Church, A.}, Kleene\index{Kleene, S.} and many others, harking back even to the 19th century ideas of Charles Babbage and Ada Lovelace\index{Lovelace, A.},\footnote{DeMol\index{DeMol, L.}, Sieg\index{Sieg, W.}, Sutner, Swade\index{Swade, D.} and Zuse.} and which was greatly clarified by Turing's notion of a computing machine, and by Post's closely related ideas, indeed depend on a fundamental \textit{discreteness} of the basic ingredients. The various very different-looking proposals for a notion of effective computability that these early 20th century logicians introduced all turned out to be \textit{equivalent} to one another, a fact that is central to our current viewpoint concerning computation, and which provides us with the \textit{Church--Turing thesis}, namely that this precise theoretical notion of ``computability'' does indeed encapsulate the idea of what we intuitively mean by an idealized ``mechanical procedure''. We find this issue discussed at some depth by numerous authors in this volume\footnote{DeMol, Sieg, Dershowitz, Sutner, Bauer and Cooper.}. For my own part, I am happy to accept the Church--Turing thesis, in this \textit{original} sense of this phrase, namely that the mathematical notion of computability---as defined by what can be achieved by Church's $\lambda$-calculus, or equivalently by a Turing machine---is indeed the appropriate ideal \textit{mathematical} notion that we require for our considerations of computability. Whether or not the universe in which we live operates in accordance with such a notion of computation is then an issue that we may speculate about, or reason about in one way or another (see, for example,~Refs.~\cite{zenil1,penrose1994}).

 Nevertheless, I can appreciate that there are other viewpoints on this, and that some would prefer to define ``computation'' in terms of what a physical object can (in principle?) achieve\footnote{Deutsch\index{Deutsch, D.}, Teuscher\index{Teuscher, C.}, Bauer\index{Bauer, A.} and Cooper\index{Cooper, B.}.}. To me, however, this begs the question, and this same question certainly \textit{remains}, whichever may be our preference concerning the use of the term ``computation''. If we prefer to use this ``physical'' definition, then all physical systems ``compute'' \textit{by definition}, and in that case we would simply need a different word for the (original Church-Turing) mathematical concept of computation, so that the profound question raised, concerning the perhaps computable nature of the laws governing the operation of the universe can be studied, and indeed questioned. Accordingly, I shall here use the term ``computation'' in this \textit{mathematical} sense, and I address this question of the computational nature of physical laws in a serious way later.

Returning, now, to the issue of the discreteness that came through the introduction of standard quantum mechanics, we find that the theory, as we understand it today, has \textit{not} developed in this fundamentally discrete direction that would have fitted in so well with our ideas of computation. The discreteness that Max Planck revealed, in 1900, in his analysis of black-body radiation (although not initially stated in this way) was in effect a discreteness of \textit{phase space}---that high-dimensional mathematical space where each spatial degree of freedom, in a many-particle system, is accompanied by a corresponding momentum degree of freedom. This is not a discreteness that could apply directly to our seemingly continuous perceptions of space and time. Nonetheless, various contributors to this volume\footnote{Bolognesi\index{Bolognesi, T.} and Lloyd\index{Lloyd, S.}.} have ventured in that more radical direction, arguing that some kind of discreteness might be revealed when we try to examine spatial separations of around the \textit{Planck length} $\textit{l}_P$ (approximately $10^{-35}$m) and temporal separations of around the Planck time (approximately $10^{-43}$s). These separations are absurdly tiny, smaller by some 20 orders of magnitude from scales of distance and time that are relevant to the processes of standard particle physics. Since these Planck scales are enormously far below anything that modern particle accelerators have been able to explore, it can be reasonably argued that a granularity in the very structure of space-time occurring at the absurdly tiny Planck scales would not have been noticed in current experiments. In addition to this, it has long been argued by some theoreticians, most notably by the distinguished and highly insightful American physicist John A. Wheeler, \cite{wheeler} that our understanding of how a \textit{quantum-gravity} theory ought to operate (according to which the principles of quantum mechanics are imposed upon Einstein's general theory of relativity) tells us that we must indeed expect that at the Planck scales of space-time, something radically new ought to appear, where the smooth space-time picture that we adopt in classical physics would have to be abandoned and something quite different should emerge at this level. Wheeler's argument---based on principles coming from conventional ideas of how Heisenberg's uncertainty principle when applied to quantum fields---involves us in having to envisage wild ``quantum fluctuations'' that would occur at the Planck scale, providing us with a picture of a seething mess of topological fluctuations. While this picture is not at all similar to that of a discrete granular space-time, it is at least supportive of the idea that something very different from a classically smooth manifold ought to be relevant to Planck-scale physics, and it \textit{might} turn out that a discrete picture is really the correct one. This is a matter that I shall need to return to later in this Foreword.

When it comes to the simulation of conventional quantum systems (not involving anything of the nature of Planck-scale physics) then, as was the case with classical systems, we find that we need to consider the smooth solutions of a (partial) differential equation---in this case the \textit{Schr\"{o}dinger equation}. Thus, just as with classical dynamics, we cannot directly apply the Church--Turing notion of computability to the evolution of a quantum system, and it seems that we are driven to look for simulations that are mere \textit{approximations} to the exact continuous evolution of Schr\"{o}dinger's wave function. Turing himself was careful to address this kind of issue, \cite{turing1948} whether it be in the classical or quantum context, and he argued, in effect, that discrete approximations when they are not good enough for some particular purpose can always be improved upon while still remaining discrete. It is indeed one of the key advantages of digital as opposed to analogue representations, that an exponential increase in the accuracy of a digital simulation can be achieved simply by incorporating additional digits. Of course, the simulation could take much longer to run when more digits are included in the approximation, but the issue here is what can \textit{in principle} be achieved by a digital simulation rather than what is practical. In theory, so the argument goes, the discrete approximations can always be increased in accuracy, so that the computational simulations of physical dynamical process can be as precise as would be desired.

Personally, I am not fully convinced by this type of argument, particularly when chaotic systems are being simulated. If we are merely asking for our simulations to represent \textit{plausible} outcomes, consistent with all the relevant physical equations, for the behaviour of some physical system under consideration, then chaotic behaviour may well not be a problem, since we would merely be interested in our simulation being realistic, not that it produces the actual outcome that will in fact come about. On the other hand, if---as in weather prediction---it is indeed required that our simulation emph{is} to provide the actual outcome of the behaviour of some specific system occurring in the world that we actually inhabit, then this is another matter altogether, and approximations may not be sufficient, so that chaotic behaviour becomes a genuinely problematic issue.\footnote{Matters relevant to this issue are to be found in \cite{wolfram,israeli,zenil2}.}

It may be noted, however, that the Schr\"{o}dinger equation, being \textit{linear}, does not, strictly speaking, have chaotic solutions. Nevertheless, there is a notion known as ``quantum chaos'', which normally refers to quantum systems that are the quantizations of chaotic classical systems. Here the issue of ``quantum chaos'' is a subtle one, and is all tied up with the question of what we normally wish to use the Schr\"{o}dinger equation \textit{for}, which has to do with the fraught issue of \textit{quantum measurement}. What we find in practice, in a general way---and I shall need to return to this issue later---is that the evolution of the Schr\"{o}dinger equation does not provide us with the unique outcome that we find to have occurred in the actual world, but with a superposition of possible alternative outcomes, with a probability value assigned to each. The situation is, in effect, no better than with chaotic systems, and again our computational simulations cannot be used to predict the actual dynamical outcome of a particular physical system. As with chaotic systems, all that our simulations give us will be alternative outcomes that are plausible ones---with probability values attached---and will not normally give us a clear prediction of the future behaviour of a particular physical system. In fact, the quantum situation is in a sense ``worse'' than with classical chaotic systems, since here the lack of predictiveness does not result from limitations on the accuracy of the computational simulations that can be carried out, but we find that even a completely precise simulation of the required solution of the Schr\"{o}dinger equation would not enable us to predict with confidence what the \textit{actual} outcome would be. The unique history that emerges, in the universe we actually experience, is but one member of the \textit{superposition} that the evolution of the Schr\"{o}dinger equation provides us with.\footnote{The question may be raised that the seeming randomness that arises in chaotic classical dynamics might be the result of a deeper quantum-level actual randomness. However, this cannot be the full story, since quantum randomness also occurs with quantized classical systems that are not chaotic. Nevertheless, one may well speculate that in the non-linear modifications of quantum mechanics that I shall be later arguing for, such a connection between chaotic behaviour and the probabilistic aspects of present-day quantum theory could well be of relevance.}

Even this ``precise simulation'' is problematic to some considerable degree. We again have the issue of discrete approximation to a fundamentally continuous mathematical model of reality.  But with quantum systems there is also an additional problem confronting precise simulation, namely the vast size of the parameter space that is needed for the Schr\"{o}dinger equation of a many-particle quantum system. This comes about because of the quantum entanglements referred to earlier. Every possible entanglement between individual particles of the system requires a separate complex-number parameter, so we require a parameter space that is exponentially large, in terms of the number of particles, and this rapidly becomes unmanageable if we are to keep track of everything that is going on. It may well be that the future development of \textit{quantum computers} would find its main application in the simulation of quantum systems. We find in this collection, some discussion of the potential of quantum computers, though there no consensus is provided as to the likely future of this interesting area of developing technology.\footnote{Schmidhuber\index{Schmidhuber, J.}, Lloyd\index{Lloyd, S.}, Zurowski\index{Zurowski, M.}.}

We see that despite the discreteness that has been introduced into physics via quantum mechanics, our present theories still require us to operate with real-number (or complex-number) functions rather than discrete ones. There are, however, proposals (e.g.  \cite{blum}) in which the notion of ``computation'' is taken in a sense in which it applies \textit{directly} to real-number operations, the real numbers that are employed in the physical theory being treated as real numbers, rather than, say, rational approximations to real numbers (such as finitely terminated binary or decimal approximations). In this way, simulations of physical processes can be carried out without resorting to approximations. This, however, can require that the initial data for a simulation be given as explicitly known functions, and that may not be realistic. Moreover, there are various \textit{different} concepts of computability with real numbers,  \cite{weihrauch,bridges,stannett,blum,pourel1} which, unlike in the situation that arose for discrete (integer-valued) variables, where the Church-Turing concept appears to have provided a \textit{single} generally accepted universal notion of ``computation'', there are many different proposals for real-number computability and no such generally accepted single version appears to be in evidence. Moreover, we unfortunately find that, according to a reasonable-looking notion of real-number computability, the action of the ordinary second-order wave operator turns out to be non-computable in certain circumstances (see~e.g.~Refs.~\cite{pourel1,pourel2}). Whatever the ultimate verdict on real-number computability might be, it appears not to have settled down to something unambiguous as yet.

There is also the question of whether an exact theory of real-number computability would have genuine relevance to how we model the physical world. Since our measurements of reality always contain some room for error---whether this be in a limit to the precision of a measurement or in a \textit{probability} that a discrete parameter might take one or another value (as sometimes is the case with quantum mechanics)---it is unclear to me how such an exact theory of real-number computability might hold advantages over our present-day (Church--Turing) discrete-computational ideas. Although the present volume does not enter into a discussion of these matters, I do indeed believe that there are significant questions of importance here that should not be left aside (for example, see  \cite{pourel2,penrose1994,weihrauch,bridges}).

Several articles in this volume address the issue of whether, in some sense, the universe actually \textit{is} a computer.\footnote{Lloyd, Deutsch, Turner\index{Turner, R} and Zuse.} To me, this seems to be a somewhat strange idea. Although I can more-or-less understand what it might mean for it to be possible to have (theoretically) a computational \textit{simulation} of all the actions of the physical universe,\footnote{Bolognesi and Szudzik\index{Szudzik, M.}.} which involves some sort of ``constructivist'' assumption\footnote{Bauer.} for the operation of the physical world, I find it much less clear what it might mean for the universe to \textit{be} a computer. Various images come to mind, maybe suggested by how one chooses to picture a modern electronic computer in operation. Our picture might perhaps consist of a number of spatially separated ``nodes'' connected to one another by a system of ``wires'', where signals of some sort travel along the wires, and some clear-cut rules operate at the nodes, concerning what output is to arise for each possible input. There also needs to be some kind of direct access to an effectively unlimited storage area (this being an essential part of the Turing-machine aspirations of such a computer-like model). However, such a discrete picture and a fixed computer geometry does not very much resemble the standard present-day models that we have of the small-scale activity of the universe we inhabit. The discreteness of this picture is perhaps a little closer to some of the tentative proposals for a discrete physical universe, such as the ``causal sets''\footnote{Bolognesi.\index{Bolognesi, T.}} that I shall briefly return to later, which represent some attempts at radical ideas for what space-time might be ``like'' at the Planck scale. 

Yet, there are some partial resemblances between such a computer-like picture and our (very well supported) present-day physical theories. These theories involve individual constituents, referred to as ``quantum particles'', where each would have a classical-level description as being spatially ``point-like''---though persisting in time, providing a classical space-time picture of a 1-dimensional ``world-line''. If these world-lines are to be thought of as the ``wires'' in the above computer-inspired picture, then the ``nodes'' could be thought of as the interaction places (or intersection points) between different particle world-lines. This would be not altogether unlike the computer image described above, though in standard theory, the topological geometry of the connections of nodes and wires would be part of the dynamics, and not fixed beforehand. Perhaps the lack of a fixed geometry of the connections would provide a picture more like the \textit{amorphous} type of computer structure also considered in this volume,\footnote{Hewitt\index{Hewitt, C.}, Teuscher\index{Teuscher, C.}, Margenstern\index{Margenstern, M.} and Wiedermann\index{Wiedermann, J.}.} than a conventional computer. However, it is still not clear how the ``direct access to an effectively potentially unlimited storage area'' is to be represented. More seriously, this is merely the \textit{classical} picture that is conjured up by our descriptions of small-scale particle activity, where the quantum ``picture'' would consist (more or less) of a \textit{superposition} of all these classical pictures, each weighted by a complex number.  Such a ``picture'' perhaps gets a little closer to the way that a \textit{quantum} computer\footnote{Schmidhuber, Margenstern, Zurowski.} might be represented, but again there are the crucial issues raised by the topology of the connections being part of the dynamics and the absence of an ``unlimited storage area'', in the physical picture, which seem to me to represent fundamental differences between our universe picture and a quantum computer. In addition to all this, there is again the matter of how one treats the \textit{continuum} in a computational way, which in quantum (field) theory is more properly the \textit{complex} rather than the real continuum. Over-riding all this is the matter of how one actually gets information out of a quantum system. This requires an analysis of the \textit{measurement problem} that I shall need to come to shortly. 

I think that, all this notwithstanding, when people refer to the universe ``being'' a computer, the image that they have is not nearly so specific as anything like that suggested above. More likely, for our ``computer universe'' they might simply have in mind that not only can the universe's actions be precisely simulated in all its aspects, but that it has \textit{no other} functional quality to it, distinct from this computational behaviour. More specifically, for our ``computer universe'' there would be likely to be some parameter $t$ (presumably a discrete one, which could be regarded as taking integer values) which is to describe the passage of time (not a very relativistic notion!), and the state of the universe at any one time (i.e. $t$-value) would have some computational description, and so could be completely encoded by a single natural number $S_t$. It would be the universe's job to compute $S_{t^\prime}$ from $S_t$ whenever $t^\prime>t$, and the universe would be considered to \textit{be} a computer provided that not only is it able always to achieve this, but---more importantly---that this is the \textit{sole} function of the universe. It seems to me if, on the other hand, the universe has any additional function, such as to assign a \textit{reality} to any aspect of this description, then it would not simply be a computer, but it would be something more than this, succeeding in providing us with some kind of \textit{ontology} that goes beyond the mere computational description.

To conclude this Foreword, I wish to present something that is much more in line with my \textit{own} views as to the relation between computation and the nature of physical reality. To begin with, I should perhaps point out that my views have evolved considerably over the decades, but without much in the way of abrupt changes. Early on I had been of a fairly firm persuasion that there should be a discrete or combinatorial basis to physics, perhaps somewhat along the lines expressed in some of the articles\footnote{Bolognesi, Schmidhuber, Lloyd, Wolfram\index{Wolfram, S.}, Zuse, Fredkin\index{Fredkin, Ed} and Zenil\index{Zenil, H.}.} in this volume. In 1967 Erwin Kronheimer\index{Kronheimer, E.} and I published a paper \cite{kronheimer,penrose1967} on the kind of causal sets referred to earlier in this Foreword, where the basic relationships between the elements are those of \textit{causality}\footnote{Bolognesi.}, mirroring the causal relations between events in continuous space-time, but where no continuity or smoothness is assumed, and where one could even envisage situations of this kind where the total number of these elements is finite.

Although I also had different reasons to be interested in spaces with a structure defined solely by causal relations---partly in view of their role in relation to singularity theorems \cite{penrose1965,hawking} (for the study of black holes and cosmology)---the causality relations not necessarily being tied to the notion of a smooth space-time manifold, I did not have much of an expectation that the true small-scale structure of our actual universe should be helpfully described in these terms. I had thought it much more probable that a different combinatorial idea, that I had been playing with a good deal earlier, namely that of \textit{spin-networks} (see~Ref.~\cite{penrose1971}) might have true relevance to the basis of physics (and indeed, much later, a version of spin-network theory was to form part of the loop-variable approach to quantum gravity, \cite{ashtekar} although the role that spin-networks acquire in loop-variable gravity is somewhat different from what I had originally envisaged).

Spin-network theory was based on one of the most striking parts of standard quantum mechanics, where a fundamental notion that is continuous in classical mechanics, is discrete in quantum mechanics, namely \textit{angular momentum} (or \textit{spin}). In fact, many of the most basic and counter-intuitive features of quantum mechanics, such as discreteness and (Bell) non-locality,\footnote{Breuer\index{Breuer, T.}, Cabello\index{Cabello, A.}, Schmidhuber\index{Schmidhuber, J.} and Zenil\index{Zenil, H.}.} are most powerfully expressed in terms of quantum-mechanical spin. The puzzling relation between the continuous array of possibilities for the direction of a spin axis in our classical space-time pictures and the discrete (or ``granular'') nature of the quantum idea of ``spin-axis direction'' had always maintained a fundamental fascination for me. This and the basic non-locality of information in quantum mechanics come together in spin-network theory, where the classical idea of a spatial ``direction'' does not arise in a well-defined way until very large spin-network structures are present in order provide a good approximation to the continuous sphere of possible spatial directions. Specific mathematical devices for calculating the often extremely complicated expressions were developed, but everything remains completely discrete, and computational in the conventional sense of the word, continuity arising only in the limit of large numbers. The need to generalize the idea of spin-networks in order that the geometry of 4-dimensional space-time might be described, rather than just the sphere of spatial directions, finally found some satisfaction in the ideas of \textit{twistor theory} (see~Refs.~\cite{penrose1967,penrose2004}, Chapter 33). This provided a different way of looking at space-time geometry from what is usual---but now the idea of discreteness underlying the basis of physics began to fade, and became superseded by the magic of complex geometry and analysis.

One normally thinks of the space-time 4-manifold as being composed of ``events'' (i.e. space-time points), which are the basic elements of the geometry. Instead, twistor theory takes its basic elements to be modelled on entire histories of massless spinning particles in free flight. By a careful combination of ideas from space-time 4-geometry and the quantum-mechanical structure of relativistic angular momentum for massless particles, the concept of ``twistor algebra'' was developed. \cite{penrose1967,penrose1971} In special relativity, the basic concept of a \textit{twistor}, which describes the kinematical structure of a spinning massless particle, finds its mathematical description as an element of the complex 4-vector space $\mathbb{T}$ referred to as ``twistor space''. The geometry of $\mathbb{T}$  relates to the real geometry of Minkowski 4-space $\mathbb{M}$ by means of an explicit geometrical correspondence, relating $\mathbb{M}$ directly to the complex 3-geometry of the \textit{projective} twistor space $\mathbb{PT}$. It turns out that the complex numbers of quantum mechanics dovetail with those of the complex geometry of twistor theory in surprising ways, and that there is an intriguing interplay between the non-locality that naturally arises in the twistor description of quantum wavefunctions and the non-locality that we actually find in quantum phenomena. \cite{penrose2005} In recent years, twistor theory has found considerable value in the calculation of high-energy scattering processes, where the rest-masses of the particles involved can be ignored, (See, for example,~Ref.~\cite{arkani}) but many of the deeper issues confronting twistor theory remain unresolved.   

It has always been an aim of twistor theory (still only partially fulfilled) that it should form a vehicle for the natural unification of quantum mechanics with general relativity. By this, I do not mean ``quantum gravity'' in the conventional sense in which this term is used. What is usually meant by quantum gravity is some scheme in which the ideas of Einstein's theory of gravity---namely \textit{general relativity} (or else perhaps some modification of Einstein's theory)---is brought under the umbrella of quantum field theory. This viewpoint is to take the laws of quantum field theory as being \textit{inviolate}, and that the ideas of general relativity must yield to those of quantum theory via some appropriate form of ``quantization''. My own view has always been different from this, as I believe that quantum theory itself, quite apart from its need to be unified with general relativity theory, is basically \textit{self-inconsistent} and that some help is needed from \textit{outside} the normal rules of quantum (field) theory. The view here is that the underlying principles of general relativity should help to supply this outside assistance.

This inconsistency is a very fundamental one, and is in a clear sense completely obvious (the ``elephant in the room''!) as we shall see. As remarked upon earlier, we take the evolution of a quantum system in isolation to be governed by the Schr\"{o}dinger equation---or, in more general terms, \textit{unitary evolution}---and for which I use the symbol ``U''. But, as was remarked upon earlier, the reality of the world that we actually observe taking place about us tends \textit{not} to be described directly by the solution $\Psi$ of this equation that we get by this U-evolution, but when an observation or ``measurement'' is deemed to have taken place, $\Psi$ is considered to ``jump'' to just one member $\Psi_r$ of a family of \textit{superposed} alternative solutions

\begin{equation}
\Psi  =   \alpha_1 \Psi_1 + \alpha_2 \Psi_2 + \ldots + \alpha_n \Psi_n
\end{equation}

where the respective squared moduli of the complex-number weightings $\alpha_1, \alpha_2, \dots, \alpha_n$, supply the respective probabilities of each $\Psi_r$ being the result (the quantities $\Psi_r$ being assumed to be all normalized and mutually orthogonal). The ``evolution process'' whereby $\Psi$ is replaced by the particular $\Psi$r that happens to come about is the \textit{reduction} of the state (collapse of the wavefunction) and I denote this process by the letter ``R''.\footnote{In Von Neumann's\index{von Neumann, J.} classic book Mathematical Foundations of Quantum Mechanics \cite{neumann}, he introduced ``R'' and ``U'' under the respective names ``process I'' and ``process II''.}

Of course, there will be many such decompositions, for a given $\Psi$, depending on the choice of \textit{basis} that is supposed to be determined by the choice of ``measuring device''. Indeed, we must allow that this measuring device is also part of the entire system under consideration, and so should have a quantum state that becomes entangled with the quantum system under examination. Nevertheless there is still taken to be a ``jump'' in the system as a whole as soon as the measurement is considered to have been made, where the different ``pointer states'' of the device are entangled with the different possible $\Psi_r$s that can result. It is \textit{obvious} that this ``jumping'' from the state of the system (consisting of both the measuring device and system under examination, together with the entire relevant surrounding environment), from before measurement to after measurement, is normally not even continuous, let alone a solution of the Schr\"{o}dinger equation: so R blatantly violates U (in almost all circumstances).

Why do physicists not normally consider this to be a contradiction in quantum mechanics? There are many responses, usually involving some subtle issue of ``interpretation'', according to which physicists try to circumvent this (seeming?) contradiction. Here is where the ``many-worlds'' viewpoint of Hugh Everett III is often invoked, \cite{everett,dewitt} whereby it is considered that \textit{all} alternative outcomes simply(!) \textit{co-exist} in quantum superposition, and that it is perhaps somehow a feature of our conscious perception processes that we always perceive only \textit{one} of these alternatives. Despite this idea's popularity among many philosophically minded physicists (or physically well-educated philosophers), I find this viewpoint very unsatisfactory. I would agree that it is indeed where we are led, if we regard the U-process as inviolate, but to me this is to be taken as a \textit{reductio ad absurdum} and a clear indication that we need to seek an improvement in current quantum mechanics. To put this another way, even if the many-worlds viewpoint is in some sense ``correct'', it is still inadequate as a description of the physical world, for the simple reason that it does not, as it stands, describe the world that we actually observe, in which we find that something extremely well approximated by the R-process \textit{actually} takes place when quantum superpositions of states that are sufficiently different from one another are involved.

What do I mean by ``sufficiently different''? It is clear that mere \textit{physical distance} apart, for the different material displacements involved in the superposed states, is not the correct criterion, because there have been well-confirmed experiments in which photon states tens of kilometres apart still maintain their quantum entanglements with one another, so that their various possible different polarization states remain in quantum superposition with each other even over such distances.\cite{tittel} However, there are reasons to expect, from various foundational principles of Einstein's general theory of relativity, (see\text{ }\cite{diosi,penrose1996,penrose2009}) that when \textit{mass displacements} between two quantum-superposed states get large, then such superpositions become unstable and ought to decay, in a roughly calculable time $\tau$, into one or the other, so that classical behaviour begins to take over from quantum behaviour. The estimate of $\tau$ is given by the formula

\begin{equation}
\tau  \approx \frac{\hbar}{E_G}
\end{equation}

where $E_G$ is the gravitational self-energy of the \textit{difference} between the mass distributions in each of two quantum states under consideration, each being assumed to constitute a stationary state if on its own. Such a decay would represent a deviation from the standard \textit{linearity} of U and might perhaps even be the result of some kind of chaotic behaviour arising in some non-linear generalization of present-day quantum mechanics. There are experiments currently under development that are aimed at testing this proposal, and we may perhaps anticipate results over the next several years (see~Ref~ \cite{marshall}).

For various reasons, partly concerned with the quantum \textit{non-locality} referred to earlier (which only begins to present substantial problems for quantum realism when R is involved, treated as a \textit{real} phenomenon), I would expect this change in current quantum mechanics to represent a \textit{major revolution} and would not be at all easy to arrive at simply by ``tinkering'' with the Schr\"{o}dinger equation. Indeed, my expectations are that such a theory would have to be \textit{non-computable} in some very subtle way. Why am I making such an assertion? The main reasons are rather convoluted, and I quite understand why some people regard my proposals as somewhat fanciful. Nevertheless, I am of the view that there is a good foundational rationale for a belief that something along these lines may actually be true!

The basic reason comes from G\"{o}del's famous incompleteness theorems, which I regard as providing a strong case for human understanding being something essentially non-computable. The central argument is a familiar one, and I still find it difficult to comprehend why so many people are unwilling to take on board what would seem to be its fairly clear implication in this regard. In simple terms, the argument can be applied to our abilities to demonstrate the truth of certain mathematical propositions---which we can take to be of the form of $\Pi_1$-\textit{sentences}. A $\Pi_1$-\textit{sentence} is an assertion that some proposed Turing computation never terminates (examples being Wiles's ``Fermat's last theorem'' and Lagrange's theorem that every natural number is the sum of four squares). We might try to encapsulate, within some algorithmic procedure $\mathbb{A}$, all possible types of argument that can, in principle, be used to establish $\Pi_1$-\textit{sentences}, according to human insight and understanding. This argument might be a \textit{proof} within some given formal system $\mathbb{F}$, where $\mathbb{A}$ would be an algorithm for checking whether a proposed proof using the rules of $\mathbb{F}$ had been correctly carried out, giving the answer YES after a finite number of steps if this is indeed the case. What the G\"{o}del(--Turing) theorem shows, in this context, is that if we have \textit{trust} in $\mathbb{A}$ (and therefore in the soundness of such an $\mathbb{F}$, with regard to $\Pi_1$-\textit{sentences}) that a ``proved'' $\Pi_1$-\textit{sentence} is indeed \textit{true} whenever $\mathbb{A}$ asserts YES, then one can explicitly exhibit a $\Pi_1$-\textit{sentence} $G$ where our trust in $\mathbb{A}$ extends also to a trust in the truth of G, even though $\mathbb{A}$ itself is shown to be incapable of directly establishing $G$. In the case of G\"{o}del's second incompleteness theorem, the $\Pi_1$-\textit{sentence} $G(=G(\mathbb{F}))$ would be an assertion of the \textit{consistency} of $\mathbb{F}$, and $G$ would be the $\Pi_1$-assertion that among the theorems of $\mathbb{F}$, there would be none whose negation is also a theorem of $\mathbb{F}$. Although our trust tells us that $\mathbb{A}$ would be unable to establish $G$ (i.e. the consistency of $\mathbb{F}$), our trust that $G$ is actually true follows from our \textit{trust} in $\mathbb{A}$ (which \textit{depends} on $\mathbb{F}$'s consistency---otherwise $\mathbb{F}$ would be able to establish $2=3$, a conclusion which we certainly would not trust). Our trust in the \textit{use} of $\mathbb{F}$ as a means of establishing the truth of $\Pi_1$-\textit{sentences} therefore carries us beyond the direct capabilities of $\mathbb{F}$, and enables us to assert that $G(\mathbb{F})$ is true, on the basis of that same trust, despite the fact that $\mathbb{F}$ does not contain G($\mathbb{F}$) among its theorems.
 
This is basically the thrust of G\"{o}del's\index{G\"odel} attack on formalism. Although the formalization of various areas of mathematics certainly has its value, allowing us to the transfer different aspects of human understanding and insight into computational procedures, G\"{o}del shows us that these explicit procedures, once known---and \textit{trusted}---cannot cover everything in mathematics that is accessible to understanding and insight.\footnote{Zizzi.\index{Zizzi, P.}} And, indeed, this applies already for the relatively limited area of $\Pi_1$-\textit{sentences}. Yet, a case can certainly be argued that this does \textit{not} yet provide a demonstration that human insight is, at root, a non-algorithmic procedure, and I list here what appear to be the main arguments in support of that case, i.e. of criticisms of the above claim that the G\"{o}del-type arguments show that human understanding is non-computational,

\begin{enumerate}
\item \textit{Errors argument}---human mathematicians make errors, so rigorous G\"{o}del-type arguments do not apply.
\item \textit{Extreme complication argument}---the algorithms governing human mathematical understanding are so vastly complicated that their G\"{o}del statements are completely beyond reach.
\item \textit{Ignorance of the algorithm argument}---we do not know the algorithmic process underlying our mathematical understanding, so we cannot construct its G\"{o}del statement.
\end{enumerate}
   
I have tried to argue elsewhere \cite{penrose1994} that (1), (2), and (3) do not invalidate the conclusion that our conscious understandings are very unlikely to be entirely the product of computational actions, and it is not my purpose to repeat such detailed arguments here. Nevertheless I briefly summarise my counter-arguments, in what follows.

The main point, with regard to (1) is that human errors are correctable. We are not so much concerned with the often erroneous gropings that mathematicians employ in their search for truth, but more the \textit{ideals} that they grope \textit{for} and, more importantly, measure their achievements against. It is their ability to \textit{perceive} these ideals that we are concerned with, if only in principle, and it is this ability to perceive ideal mathematical truth that we are concerned with here, not the errors that we all make from time to time. (It may be evident from these comments that I \textit{do} regard mathematical truth---especially with regard to matters so straight-forward as $\Pi_1$-\textit{sentences}---as something absolute, and external to ourselves. But I appreciate that others\footnote{DeMol\index{DeMol, L.}.} are sometimes less sympathetic to this kind of viewpoint. I do not believe, however, that one's philosophical standpoint in this respect significantly affects the arguments that I am putting forward here.) With regard to (2) the point is somewhat similar. If the algorithms were \textit{in principle} to be known, then their size or complication is of no real concern. This applies to a great many mathematical arguments. In Euclid's proof of the infinity of primes, for example, we need to consider primes that are so large that there would be no way to write them down explicitly in the entire observable universe, and to calculate the product of them all up to some such size in practice, is even more out of the question. But all this is irrelevant for the proof. Similar points apply to (2).

\setcounter{footnote}{0}

The argument (3) is, however, much more relevant to the discussion, and was basically G\"{o}del's own reservation (referred to in the commentaries here\footnote{Sieg.\index{Sieg, W.}}) with regard to making the strong conclusion that I am arguing for here. Rather than pushing the logical argument further, which is certainly possible to do (see~Ref.~\cite{penrose1994}). I shall here merely indicate the extraordinary improbability of the needed algorithmic action arising in our heads, by the process of \textit{natural selection}. Such an algorithm would have to have extraordinary sophistication, so as to be able to encapsulate, in its effective ``formal system'' many steps of ``G\"{o}delization''. As an example, I have pointed out elsewhere \cite{penrose1998} that whereas \textit{Goodstein's theorem}\index{Goodstein's theorem}, \cite{goodstein} whose meaning\footnote{Velupillai.} is easily accessible even to those with little mathematical knowledge other than basic numerical notation, has been shown by Kirby and Paris \cite{kirby} to be inaccessible by first-order Peano arithmetic\index{Peano arithmetic} (without a ``G\"{o}delization''\index{G\"{o}delization} step, that is), yet this theorem can be readily seen to be true through mathematical understanding. If our mathematical understanding is achieved by some (unknowable, but sound) algorithmic procedure, it would be a total mystery how it could have arisen through natural selection, when the experiences of our remote ancestors could have gained no benefit whatsoever from having such a sophisticated yet totally irrelevant algorithm planted in their brains!

If, then, it is accepted that our understanding of mathematics is not an algorithmic process, we must ask the question what kind of process can it be? A key issue, it seems to me, is that genuine \textit{understanding} (at least in our normal sense of this word) is something that requires awareness---as it would seem to me to be a misuse of the word ``understanding'' if it could be genuinely applied to an entity that had no actual \textit{awareness} of the matter under its consideration. Awareness is the passive form of consciousness, so it seems to me that it was the evolutionary development of consciousness that is the key, and that such a quality could certainly have come about through natural selection, being able to confer an enormous selective advantage on those creatures possessing it. In saying this, I am expressing the view that consciousness is indeed \textit{functional} and is not an ``epiphenomenon'' that simply happens to accompany certain kinds of cognitive processes. This view is certainly an implication of the quality of ``understanding'' requiring conscious awareness, since understanding is certainly functional. 

I should make clear that I am making no claim to know---or to be able to define---what consciousness actually \textit{is}, but its role in underlying ``understanding'' (whatever \textit{that} is) seems to me to be of great evolutionary value, and could readily arise as a product of natural selection. I should also make clear that I am regarding the consciousness issue as a \textit{scientific} one, and that I do \textit{not} take the view that these are matters that are inaccessible to scientific investigation. I also take it that healthy wakeful human brains (as well as whatever other kind of animal brains may turn out also to be similarly capable) are able, somehow, to evoke consciousness by the application of those very same physical laws that are present throughout the universe, even though consciousness itself comes about only in the \textit{very} special circumstances of organization that are needed to promote its appearance.

What kind of circumstance could that be, if we are asking for some sort of non-computable action to come about---when we bear in mind that the deterministic differential equations of classical or quantum physics seem to be of an essentially computable nature? My response to this query is that the non-computability must lie in hitherto undiscovered laws that could be of relevance here. (I am ignoring the issue, referred to earlier, of the discrete computational simulation of a continuous evolution. Yet, I do accept that there might be some questions of genuine relevance here that ought to be followed up more fully.) As far as I can see, the only big unknown, in physical laws, that could have genuine relevance here, is the U/R puzzle of quantum mechanics, referred to above. In almost all processes that take place, we have no need of the presumed New Theory that is to go beyond current quantum mechanics, mainly because its effects would go un-noticed, being swamped by the multifarious random influences of environmental decoherence. But, in the brain, there might be relevant structures able to preserve quantum coherence up to a length of time at which the previously mentioned $\tau  \approx \hbar/E_G$ criterion actually becomes relevant. Then, the normal purely probabilistic action that standard quantum theory's R-process provides us with is to be replaced by some subtle non-computational decision as to which choice the state reduction leads to. With a sophisticated brain organization, where the synaptic responses are sensitive to these choices, we can imagine that the output of the brain could indeed be usefully non-computational. This, indeed, is the basis of the ``orchestrated objective reduction'' (\textbf{Orch-OR}) scheme that Stuart Hameroff and I have proposed some years ago, where the above ``relevant structures'' would be \textit{neuronal microtubules} of the appropriate type (see~Refs.~\cite{hameroff,penrose1996,penrose2011}).

It is hardly surprising that such a proposal has met with some considerable scepticism, mainly for the very understandable reason that to have body-temperature quantum coherence at anything like the level required is enormously far beyond the expectations of standard physical calculations applied to simplified models of cells. \cite{tegmark} Nevertheless, biological cells are, in fact, highly sophisticated structures,\footnote{Margenstern\index{Margenstern, M.}, Ehrenfeucht\index{Ehrenfeucht, A.} et al., Rozenberg\index{Rozenberg, G.} and Zenil\index{Zenil, H.}.} and one may reasonably expect that, when the structures of certain cell parts are dedicated in the appropriate directions, their behaviour might exhibit quite unusual quantum-mechanical properties.\footnote{Zizzi\index{Zizzi, P.}.} In fact, recent experiments carried out in Japan by Anirban Bandyopadhyay \cite{bandyopadhyay} and his co-workers appear to have demonstrated that highly intriguing quantum-coherent effects \textit{do} actually take place in body-temperature neuronal microtubules. These results are, as of now, preliminary, but they do appear to provide some encouragement for the \textbf{Orch-OR} scheme, and it will be very interesting to see how things develop. 

Even if all of this is accepted, we may still ask what would be the use of a little bit of non-computable action, from time to time, for the operation of the brain? Indeed, there would not be much value in this unless the quantum coherence is of a very global character, involving large areas of the brain, and the process would have to act in some globally coherent way. This is indeed the \textbf{Orch-OR} picture, and we take it that moments of consciousness occur when state reduction occurs at many sites (in microtubules) at once in an \textit{orchestrated} way, so that the synapse strengths are influenced in many places and a concerted influence results, as would be expected for conscious actions. The results of particular acts of conscious understanding would be unlikely to be usually anything simple, and would depend upon the experience of memories as well as on logic. But the non-computable ingredient is taken to be essential, for the reasons described above. According to this view, our conscious actions are calling upon parts of physics---encompassed in a New Theory that is presently unknown in detail. The impact of this theory on processes \textit{not} organized in this way would not be evident. But it would make its mark on systems---such as wakeful healthy human brains---where it emerges as conscious actions and perceptions. The non-computable effects of this New Theory would emerge in this way and result in actions that are described as ``hypercomputational''.

How far outside the normal scheme of computational physics would these hypercomputational actions be? Since the G\"{o}delian insight that allows us to transcend a given trusted formal system $\mathbb{F}$ provides this insight in the form of a $\Pi_1$-\textit{sentence}, namely $G(\mathbb{F})$, we might expect that we could model such hypercomputational actions in the form of a Turing \textit{oracle-machine},\footnote{Chaitin, Dershowitz.}
where the oracle is able to assert the truth or falsity of $\Pi_1$-\textit{sentence}. However this would not be sufficient (nor does it appear to be necessary), as we can apply a G\"{o}del-type ``diagonalization''\index{diagonalization method} insight again on $\Pi_1$-\textit{sentence}-oracle\index{oracle machines} machines to transcend these devices also. In a recent article, \cite{penrose2012} I consider a type of oracle that I refer to as a ``cautious oracle'', which is intended to model a little more closely the kind of thing that one might consider idealized human mathematicians might be capable of, where the cautious oracle can examine a $\Pi_n$-\textit{sentence} (for any natural number $n$) and either respond ``true'' or ``false'' (necessarily truthfully in each case), or else confess to being unable to supply an answer or, failing any of these, simply continue pondering indefinitely without ever providing an answer at all. Again a G\"{o}del-type diagonalization allows us the insight to transcend any such a device's capabilities! Whatever kind of hypercomputational\index{hypercomputation} capabilities such a ``New Theory'' might confer, it appears to be something very subtle. It is some sort of never-ending capability of being able to ``stand back'' and contemplate whatever structure had been considered previously. This seems to be a quality that consciousness is able to achieve, but how one incorporates this kind of thing into a physical theory is hard to imagine, as our present-day theories stand.

\bibliographystyle{ws-rv-van}

\end{document}